\documentclass[aps,twocolumn,superscriptaddress]{revtex4-1}

\usepackage{amsmath,amssymb,amsfonts,bm,color,graphicx,tabularx}

\usepackage[unicode=true,colorlinks=true]{hyperref}

\usepackage[etex=true,export]{adjustbox}

\usepackage{ulem}

\hypersetup{linkcolor=blue,citecolor=blue,urlcolor=blue}

\renewcommand{\v}[1]{{\bf #1}}

\newcommand{\beq}{\begin{equation}}
\newcommand{\eeq}{\end{equation}}
\newcommand{\beql}{\begin{equation*}}
\newcommand{\eeql}{\end{equation*}}
\newcommand{\beqn}{\begin{eqnarray}}
\newcommand{\eeqn}{\end{eqnarray}}

\begin{document}

\title{Superconducting spin properties of Majorana nanowires and the associated superconducting anomalous Hall effect}
\author{Li Chen}
\author{Ying-Hai Wu}
\email{yinghaiwu88@hust.edu.cn}
\author{Xin Liu}
\email{phyliuxin@hust.edu.cn}
\affiliation{School of Physics and Wuhan National High Magnetic Field Center, Huazhong University of
Science and Technology, Wuhan, Hubei 430074, China}
\date{\today}

\begin{abstract}
It is difficult to unambiguously confirm the existence of Majorana zero modes (MZMs) due to the absence of smoking-gun signatures in charge transport measurements. Recent studies suggest that the spin degree of freedom of MZMs may provide an alternative detection method. We study the spin properties of the superconducting state in Majorana nanowires and the associated unconventional Josephson effect with realistic experimental parameters taken from [Zhang {\it et al.}, \href{https://www.nature.com/articles/nature26142}{Nature 556, 74 (2018)}]. For a superconducting thin film with in-plane polarized spin-triplet pairing, an out-of-plane electric field can generate a supercurrent perpendicular to both the superconducting spin polarization and the electric field, so we name this phenomena as superconducting anomalous Hall effect (ScAHE). In a Majorana nanowire, the regime with finite polarized spin-triplet pairing almost coincides with the chiral regime, which includes the topological regime. We further study the effects of polarized spin-triplet pairing in two types of Josephson junctions. One dramatic finding is that SOC can induce an anomalous supercurrent at zero phase difference only in the U-shape junction, a basic ingredient of scalable topological quantum computation. This can be viewed as a consequence of the ScAHE. Our work reveals that the spin degree of freedom is indeed helpful for detecting MZMs.
\end{abstract}

\maketitle

\section{Introduction}
The essential ingredient of topological quantum computation (TQC)~\cite{Preskill2004,Nayak2008,Pachos2012} is exotic emergent particles that obey non-Abelian braiding statistics~\cite{Kitaev2001,Kitaev2006}. Majorana zero modes (MZMs) in superconductor-semiconductor hybrid nanowires with strong spin-orbit coupling (SOC) and magnetization, refereed to as Majorana nanowire, has arisen~\cite{Sau2010,Lutchyn2010,Oreg2010,Sau2010a} as the most promising candidate after the great experimental progresses~\cite{Mourik2012,Deng2012,Rokhinson2012,Das2012,Wang2012,Churchill2013,Xu2014,Nadj-Perge2014,Chang2015,Albrecht2016,Zhang2018}. The unambiguous confirmation of MZMs turns out to be very challenging even in the best available Majorana nanowires. The presence of zero-bias electrical conductance peak \cite{Sengupta2001}, which was proposed as a strong signature of MZMs \cite{Law2009,Flensberg2010,Pikulin2012}, cannot be taken as irrefutable evidence because trivial Andreev bound states (ABSs) may also results in robust and quantized zero-bias peaks~\cite{Liuchunxiao2017,Moore2018}. The Josephson coupling between two MZMs lead to an unusual Josephson effect with 4$\pi$-periodic Josephson current~\cite{Kitaev2001,Fu2009,Lutchyn2010,Wiedenmann2016,Bocquillon2016,Laroche2017}, but such a phenomenon may again be attributed to the presence of trivial ABSs~\cite{Chiu2018}. The inability of pinning down MZMs using zero-bias peak and fractional Josephson effect motivates us to search for other experimental probes that can clearly distinguish MZMs and other states. 

In recent works~\cite{He2014,Liu2015}, it has been shown that, for time-reversal symmetry breaking topological superconductors, the superconducting correlations of MZMs are fully spin-polarized. This unique property has attracted more and more attention because it suggests that the spin degree of freedom of MZMs can provide an invaluable alternative method for their detection and manipulation. For conductance measurement, spin selective Andreev reflection at zero bias has been predicted theoretically~\cite{He2014} and observed experimentally in the vortex core of superconducting proximitized surface of three-dimensional topological insulators~\cite{Sun2016} and the ferromagnetic atomic chain \cite{Jeon2017}. For Josephson effect, SOC-tunable fractional Josephson effect has been studied in ideal one-dimensional Majorana nanowires~\cite{Liu2016}. 

\begin{figure}[tb]
\centering
\includegraphics[width=1.0\columnwidth]{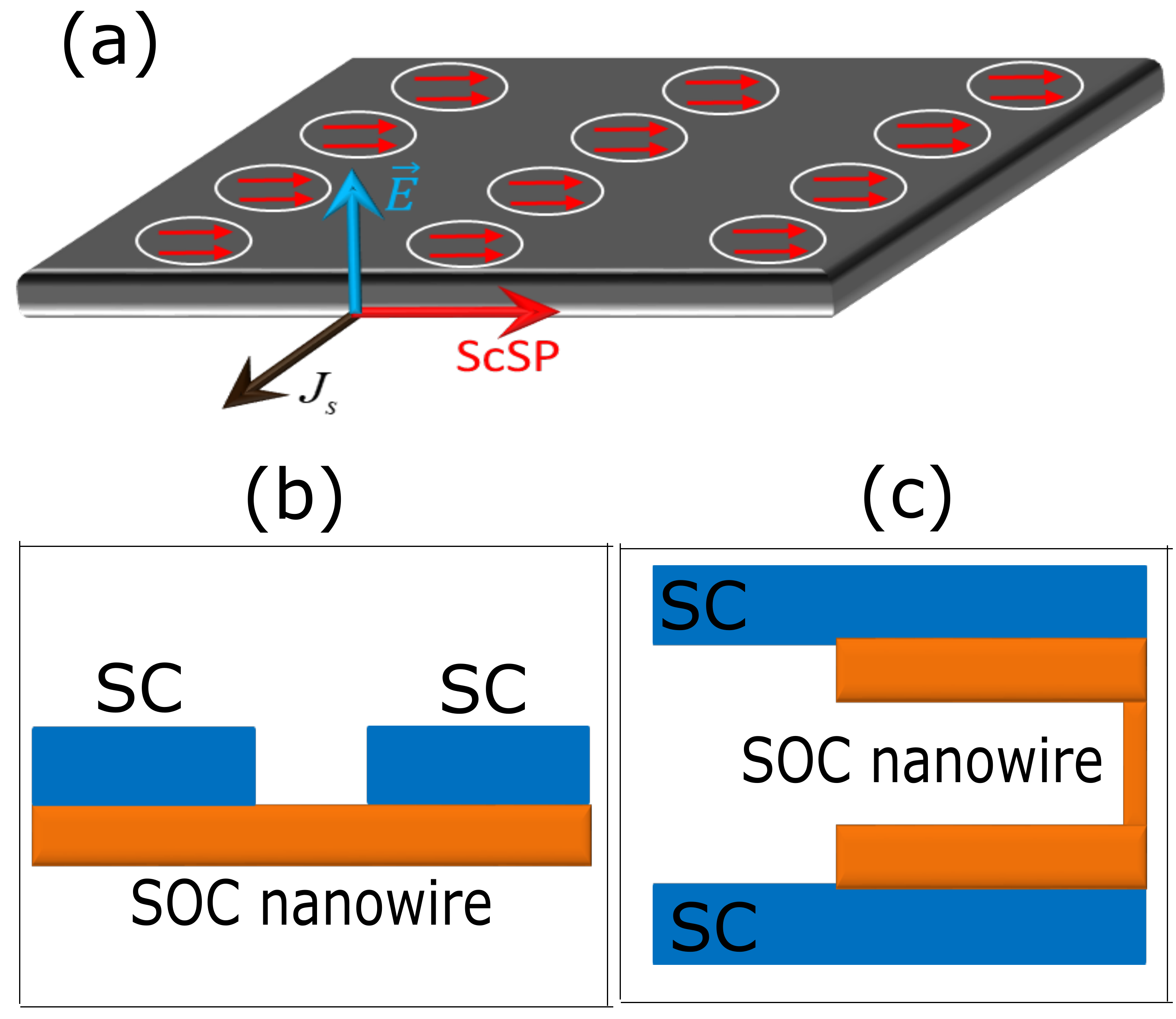}
\caption{(a) Superconducting anomalous Hall effect. The white elliptic with two red arrows indicating Cooper pairs with ScSP along $x$ direction. The blue, black and red arrows indicate the directions of external electric field $\bm{E}$, supercurrent $J_s$ and ScSP respectively. (b) Schematics of the common-shape Josephson junction. (c) Schematics of the U-shape Josephson junction.}
\label{device}
\end{figure}

To make contact with experiments, the false signals caused by inevitable trivial states in actual systems should be carefully eliminated. In this work, the superconducting spin properties of quasi-1D Majorana nanowires are thoroughly investigated, which include the contributions of not only MZMs but also trivial ABSs and bulk states. To better understand the spin properties of superconductors and the application in Majorana physics, we begin with a very general discussion of the superconducting spin polarization (ScSP), which characterizes the phenomenon of electrons with one particular spin having stronger superconducting correlations than those with the opposite spin~\cite{Leggett1975}, in a thin film with Rashba SOC. We found that the Rashba SOC only lead to spin-dependent anomalous velocity for spin-triplet Cooper pairs, which can be interpreted as the counterpart of what it does to individual electron in anomalous Hall effect \cite{Nagaosa2009} and spin hall effect \cite{Murakami2003,Sinova2004}. It is obvious that this additional anomalous velocity should be absent for spin-singlet Cooper pairs since their spin is zero. When the ScSP is finite, the spin-dependent anomalous velocity of Cooper pairs can lead to a supercurrent, which is perpendicular to the ScSP, even in the absence of phase gradient in the superconducting thin film [Fig.~\ref{device}(a)]. This phenomenon has the same origin as the anomalous Hall effect in non-interacting electrons, so we name it as superconducting anomalous Hall effect (ScAHE). To the best of our knowledge, it has not been studied in previous works.

With ScSP and ScAHE kept in mind, we then focus on Majorana nanowires with system parameters adopted from a recent experiment~\cite{Zhang2018}. The ScSP of the bulk condensates below the superconducting gap is found to show a sharp peak when the chemical potential in the chiral regime, where there are odd number of electron bands at Fermi surface. The chiral regime is the prerequisites of realizing topological superconductivity in Majorana nanowire. We further study the ScSP of the subgap states confined in the common-shape [Fig.~\ref{device}(b)] and U-shape [Fig.~\ref{device}(c)] Josephson junctions. In both cases, the ScSP exhibit similar behaviors as for the bulk condensate. However, the Josephson currents in these two settings are very different. A finite Josephson current at $\phi=0$ only occurs in the U-shape junction, which is consistent with the ScAHE discussed before. This current has a sharp peak in the topologically non-trivial regime but decays rapidly to zero in the trivial regime even in the presence of trivial ABSs. We thus propose that the observation of ScAHE in the U-shape junction, in addition to the zero-bias conductance peak and the $4\pi$ Josephson effect, would provide strong support for the existence of MZMs. As the U-shape junction is one basic building block of the scalable Majorana qubit architecture~\cite{Karzig2017}, our study is also important for the next step towards TQC.

This paper is organized as follows. In Sec.~\ref{Section2}, we first briefly review the theory of ScSP and then study the ScAHE for general superconducting thin film. In Sec.~\ref{Section3}, we study the ScSP of the bulk superconducting condensates in both strictly 1D and quasi-1D nanowires and the ScSP of the subgap bound states in Josephson junctions. In Sec.~\ref{Section4}, we demonstrate that the SOC-induced Josephson current can only occur in the U-shape junction, which can be used to confirm the presence of ScAHE. We further calculate the energy-phase relation of the subgap bound states and the associated Josephson current in the U-shape junction. In Sec.~\ref{Section5}, we conclude with a discussion about the relation between our work and previous ones. 

\section{Spin property of Cooper pairs}
\label{Section2}

Before performing detailed calculations, we give a brief introduction of ScSP and the associated supercurrent. For a superconductor in equilibrium described by the BdG Hamiltonian, the superconducting condensates of each eigenstate can be grouped as a $2{\times}2$ matrix~\cite{Liu2015} (up to a normalization constant)
\beqn\label{d}
\widehat{\Phi}(n,\bm{r}) &=& \int_{-\infty}^{\infty} {\rm Im} \left[ \mathcal{F}^{\rm R}(n,\bm{r})-\mathcal{F}^{\rm A}(n,\bm{r}) \right] dE \nonumber \\
&=& \left[\begin{array}{cc}
\Psi_{\uparrow \downarrow} (n,\bm{r}) & \Psi_{\uparrow \uparrow} (n,\bm{r}) \\
-\Psi_{\downarrow \downarrow} (n,\bm{r}) & \Psi_{\downarrow\uparrow} (n,\bm{r}) 
\end{array}\right] \nonumber \\
&=& d_0(n,\bm{r}) \sigma_0 + \bm{d}(n,\bm{r})\cdot \bm{\sigma},
\eeqn
with 
\beql
\Psi_{\sigma,\sigma{'}}(n,\bm{r})=\psi_{e,\sigma}(n,\bm{r}) \psi^{*}_{h,\sigma\rq{}}(n,\bm{r}).
\eeql
The subscript $n$ indicates the $n$-th eigenstate of the BdG Hamiltonian in the basis $({c}_{\uparrow}, {c}_{\downarrow}, {c}_{\downarrow}^{\dag}, -{c}_{\uparrow}^{\dag})^T$, the position $\bm{r}$ is the center of mass coordinates, $\mathcal{F}^{\rm R(A)}$ is the anomalous retarded (advanced) Green's function, $\sigma_{0}$ and $\bm{\sigma}$ are the identity matrix and Pauli matrices respectively, $\bm{d}_0$ and $\bm{d}$ are the spin-singlet and spin-triplet pairing amplitudes respectively, and $\psi_{e,\sigma}$ and $\psi_{h,\sigma}$ represent the electron and hole components of the eigenstate. For a translationally invariant system, the index $n$ can be replaced by momentum $\bm{k}_i$ with $i$ denoting the $i$-th eigenstate with momentum $\bm{k}$. The ScSP is defined as \cite{Leggett1975} 
\beqn\label{ScSPs}
\bm{S}(n,\bm{r}) = i \left[ \bm{d}(n,\bm{r}) \times \bm{d}^*(n,\bm{r}) \right].
\eeqn
whose $z$-component reads
\beqn\label{ScSP-1}
&& S_z(n,\bm{r}) = |\Psi_{\uparrow\uparrow}(n,\bm{r})|^2-|\Psi_{\downarrow\downarrow}(n,\bm{r})|^2,
\eeqn
One can quantify the difference of superconducting correlations between the electrons in spin-up and spin-down bands using
\beqn
&& S_z(\bm{r}) = \sum_{n} S_z(n,\bm{r}) f(E_n),
\eeqn
where $f(E_n)=1/(1+\exp(E_n/kT))$ is the Fermi distribution function with $k$ the Boltzmann constant and $T$ the temperature. This quantity vanishes in spin-singlet superconductors in which the paired electrons always have opposite spins. It should als been emphasized that $S_z(\bm{r})$ is completely different from the spin polarization of electrons given by
\beqn
\rho_{\uparrow}-\rho_{\downarrow} = \int_{-\infty}^{\infty}{\rm Im} \left( g^{\rm R}_{\uparrow \uparrow} - g^{\rm A}_{\downarrow\downarrow}\right) f(E) dE,
\eeqn
with $g^{\rm R}$ and $g^{\rm A}$ being the retarded and advanced Green's function of electrons respectively~\cite{Leggett1975}.

When electromagnetic potential or Rashba SOC is added to the system, the kinetic momentum operator of electrons changes to $\bm{\widehat{p}}-e\bm{A}$ or $\bm{\widehat{p}}-\lambda \bm{\sigma}\times\bm{\nabla}V$ respectively ($\bm{A}$ is the vector poential , $V$ is the scalar potential, $\lambda$ is the SOC strength). As a consequence, the eigenstates of the system are transformed as
\beqn
\psi(\bm{A})=\widehat{U}(A) \psi(\bm{A}=0) \;\; \text{and} \;\; \psi(\lambda)=\widehat{U}(\lambda) \psi(\lambda=0) \nonumber
\eeqn
where the unitary operators
\beqn\label{trans}
&&\widehat{U}(\bm{A})=\exp \left( -\frac{i}{\hbar} \int_0^{\bm r} e\bm{A} \sigma_0 \tau_z \cdot \bm{dr} \right), \nonumber \\
&&\widehat{U}(\lambda)=\exp \left[ -\frac{i}{\hbar} \int_0^{\bm r} \lambda (\bm{\sigma} \times \bm{\nabla} V) \tau_0 \cdot \bm{dr} \right].
\eeqn
with Pauli matrices $\bm{\tau}$ acting in the particle-hole space. For all types of superconductors, the superconducting condensate in the presence of a vector potential satisfies
\beqn
\widehat{\Phi}(\bm{A})= \exp \left(-\frac{i}{\hbar}\int_0^{\bm r} 2e\bm{A} \sigma_0 \cdot \bm{dr}\right) \widehat{\Phi}(\bm{A}=0),
\eeqn
where only the identity matrix appears because $\bm{A}$ couples to the charge degree of freedom and does not distinguish the two spin directions. This is equivalent to saying that the kinetic momentum for all types of the superconducting condensates is modified to be $\bm{p}-2e\bm{A}$, which should be expected since the Ginzburg-Landau theory also yield this result. The situation gets more complicated in the presence of SOC. Here, we consider a superconducting thin film in the $x-y$ plane and apply an out-of-plane electric field (which cannot lead to any AC supercurrent) to generate the Rashba SOC, 
\beql
\widehat{H}_{so}=\lambda E_z (p_x \sigma_y-p_y \sigma_x)\tau_z.
\eeql
The eigenstates of the system with $\lambda{\neq}0$ can be obtained from those at $\lambda=0$ via the gauge transformation
\beqn
\psi(\lambda)&=& \widehat{U}(\lambda)\psi(\lambda= 0), \nonumber \\
\widehat{U}(\lambda)&=& \exp[-i\lambda E_{z}\tau_{0}(x\sigma_{y}-y\sigma_{x})]. \nonumber
\eeqn
as indicated by Eq.~\eqref{trans}. It is easy to verify that the kinetic momentum of the spin-singlet superconducting condensate will not be affected by this transformation. However, the kinetic momenta of the spin-triplet superconducting condensate is changed to (up to the first order of $E_z$)
\beqn
\bm{d}^{\alpha} \cdot \bm{\sigma} &=& \widehat{U}(\lambda) \bm{d}^{0} \cdot \bm{\sigma} \widehat{U}^{\dagger}(\lambda)\nonumber \\
&\approx& \exp(-2i\lambda E_z x \sigma_y) d_x^0 \sigma_x + \exp(2i\lambda E_z y \sigma_x) d_y^0 \sigma_y \nonumber \\
&& + \exp[-2i\lambda E_{z} (x\sigma_{y} - y \sigma_{x})] d_z^0 \sigma_z,
\eeqn
with $\bm{d}^{\alpha}$ ($\bm{d}^0$) being the $d$-vector in the presence (absence) of SOC. As shown in Table.~\ref{table:table1}, additional terms induced by SOC appear in the kinetic momenta for the superconducting condensates of different spin configurations.

\begin{table}
\begin{center}
\begin{adjustbox}{max width=0.5\textwidth}
\begin{tabular}{|c|c|c|}
  \hline
  pairing & K-momentum (x) & K-momentum (y) \\
  \hline
  $d_0\sigma_0$ & $\widehat{p}_x$ & $\widehat{p}_y$\\
  \hline
  $d_x\sigma_x$ &$\widehat{p}_x- 2\lambda E_z \sigma_y$ & $ \widehat{p}_y$ \\
  \hline
  $d_y\sigma_y$ & $\widehat{p}_x$ & $ \widehat{p}_y + 2\lambda E_z \sigma_x $\\
  \hline
  $d_z\sigma_z$ & $\widehat{p}_x- 2\lambda E_z \sigma_y$  & $\widehat{p}_y + 2\lambda E_z\sigma_x $ \\
  \hline
\end{tabular}
\end{adjustbox}
\caption{The K-momentum for four pairings in superconducting thin film with Rashba spin-orbit coupling.}
\label{table:table1}
\end{center}
\end{table}

The existence of SOC modifies kinetic momenta and thus can give rise to unconventional supercurrent. This effect is more more pronounced in a Josephson junction with zero phase difference. In this case, the superconducting condensates must satisfy $\nabla\Psi=0$ to guarantee that the wave functions are single-valued, which is valid even in the presence of SOC. The supercurrent is solely determined by the additional kinetic momenta term induced by SOC (see Table~\ref{table:table1}) as (see appendix A for the detailed proof)
\beqn\label{sshe}
J_{i}^{s}=\lambda E_z \epsilon_{ij} S_j,
\eeqn
One can see that the supercurrent, the ScSP, and the external electric field are mutually perpendicular to each other. For this reason, we refer to this phenomenon as the ScAHE. It is obvious that Eq.~\eqref{sshe} is invariant under time-reversal operation and allows for dissipationless transport. For a Josephson junction made from superconductors with intrinsic ScSP (e.g. the A1 phase of $^3$He), SOC in the normal regime will induce a spin-triplet Josephson current through the junction even if there is no flux in the superconducting circuit. It is unfortunate that spin-triplet superconductors are very rare in nature, but we shall demonstrate below that ScSP can also appear in ordinary $s$-wave superconductors supplemented with SOC and magnetization, including Majorana nanowires~\cite{Liu2015}. This work not only identifies possible platforms for experimental observation of ScSP, but also provides a new method of probing MZMs.

\begin{figure}[htb]
\centering
\includegraphics[width=1\columnwidth]{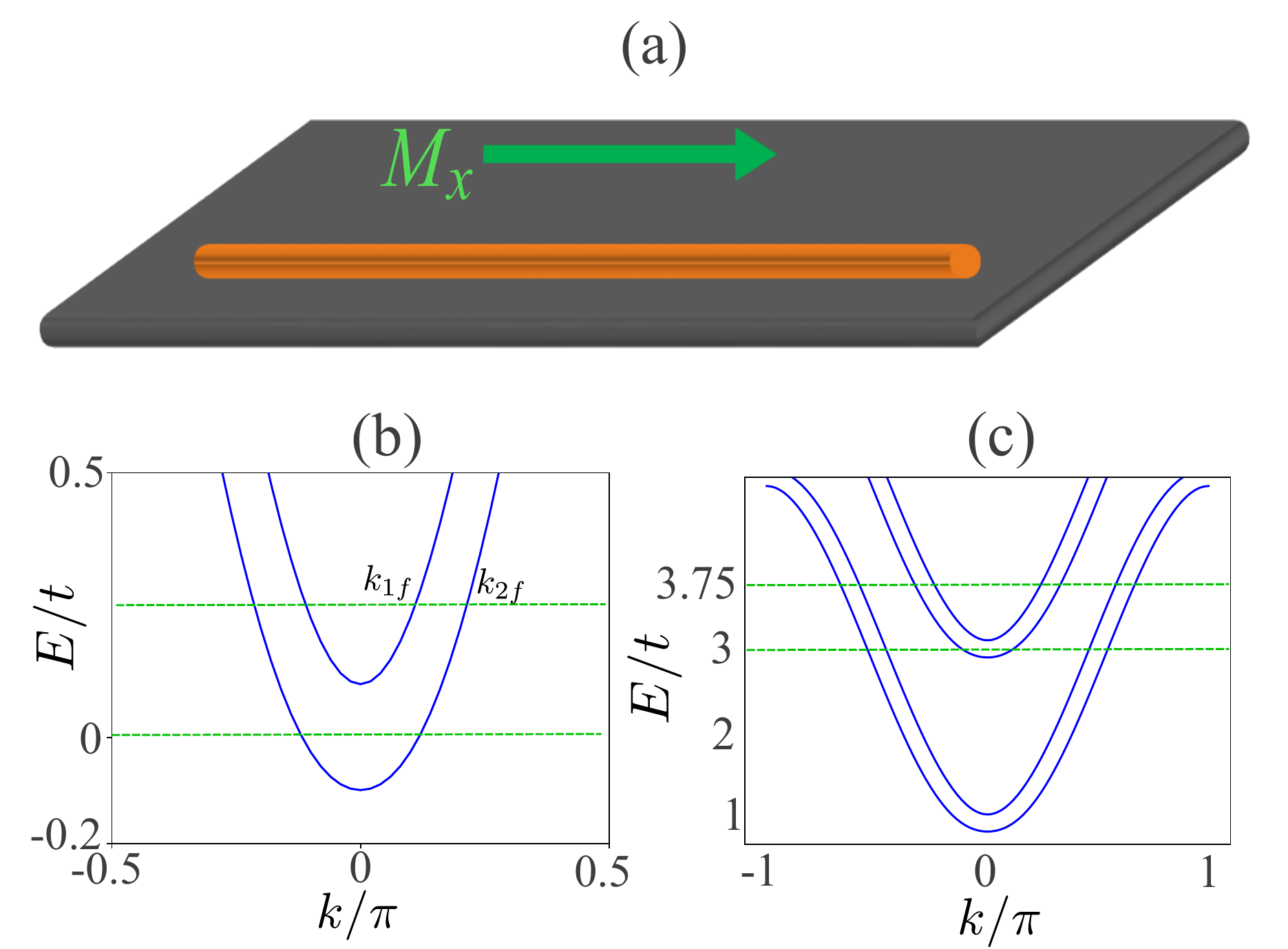}
\caption{(a) Majorana nanowire(orange) with applied magnetic field (green arrow) along the nanowire. (b)(c)  The electron dispersion in the nanowire with one and two subbands respectively. The lower and upper green dashed lines indicate odd and even number of bands across the Fermi surface which correspond to the chiral and non-chiral regimes respectively.}
\label{band}
\end{figure}

\section{ScSP in Majorana nanowires}
\label{Section3}

For a Majorana nanowire [Fig.\ref{band}(a)], ScSP may be generated by two different types of states: bulk states whose energies are not in the superconducting gap and subgap states including trivial ABSs and Majorana bound states. Their properties are studied separately in the following two subsections.

\subsection{Bulk states}

\begin{figure*}[htb]
\centering
\includegraphics[width=2.\columnwidth]{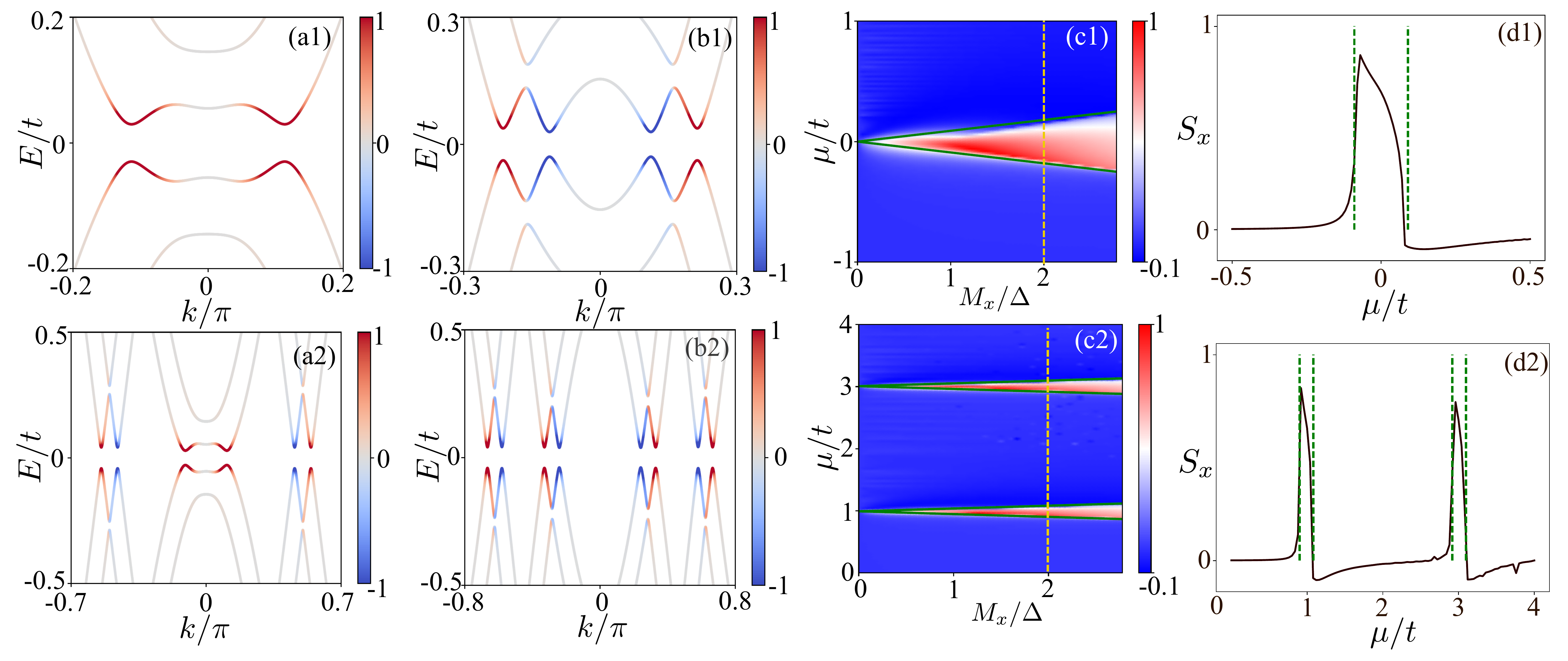}
\caption{ The upper and lower panels correspond to real-1D and quasi-1D MNWs respectively. The red (blue) color indicate the positive (negative) ScSP. (a1,a2) The spectrum of the BdG Hamiltonian of MNW with the chemical potential in the chiral regime. (b1,b2) The spectrum of the BdG Hamiltonian of MNW with the chemical potential in the non-chiral regime. (c1,c2) The net ScSP as a function of $\mu$ and $M_x$. (d1,d2) The net ScSP as a function of $\mu$ along the line cut (dashed lines in (c1,c2)).}
\label{spin-bulk-state}
\end{figure*}

To understand the ScSP of bulk states, periodic boundary condition is used in the Majorana nanowire so there is no complication due to subgap states. As a relatively simple starting point, we consider a strictly 1D system along the $x$ direction described by the continuous Hamiltonian
 \beqn\label{Ham-1}
H_{1D}(k) &=& \left( \frac{\hbar^2k^2}{2m^*}-\mu \right) \tau_z\otimes\sigma_0-M_x \tau_0\otimes \sigma_x \nonumber \\
&+& \alpha k \tau_z\otimes \sigma_y+\Delta \tau_x\otimes\sigma_0,
\eeqn
where $m^*$ is the effective mass of electrons, $\mu$ is the chemical potential, $M_x$ the Zeeman coupling strength, $\alpha{\propto}E_z$ is the SOC strength, and $\Delta$ is the proximity induced superconducting gap. Motiaved by the experimental results in the Ref.~\onlinecite{Zhang2018}, we choose the following representative system parameters: $m^*=0.02m_e$, $\alpha=2.5$meV, $\Delta=0.9$meV. After some straightforward calculations using Eq.~\eqref{d}, the superconducting condensates for various spin states are found to be
\beqn\label{d-vector}
d_{0} &\propto&  \Delta ^2+\alpha^2 k^2+\epsilon _k^2 - E^2-M_x^2, \nonumber \\
d_{x} &\propto&  2EM_x, \ \ d_{y} \propto 2\alpha k \epsilon_k \ \ d_{z} \propto i2\alpha k M_x, \nonumber \\
\epsilon_k&=& \frac{\hbar^2 k^2}{2m^*} - \mu = \frac{\hbar^2 (k^2-k_{f}^2)}{2m^*}, \ \ k_{f} = \sqrt{2m^*\mu}/\hbar.
\eeqn

Since inversion symmetry is broken in our system, the superconductor has both spin-singlet and spin-triplet pairings even though the superconducting gap function is spin-singlet. Accordingly, the ScSPs calculated based on Eq.~\eqref{ScSPs} take the form
\beqn\label{Spin-pl}
S_x(k) &=& |\langle{c_{-k\rightarrow}c_{k\rightarrow}}\rangle|^2-|\langle{c_{-k\leftarrow}c_{k\leftarrow}}\rangle|^2=-4\alpha^2 k^2 M\epsilon_k,\nonumber \\
S_y(k) &=& |\langle{c_{-k\nearrow}c_{k\nearrow}}\rangle|^2-|\langle{c_{-k\swarrow}c_{k\swarrow}}\rangle|^2\propto-4\alpha k E M^2, \nonumber \\
S_z(k) &=& |\langle{c_{-k\uparrow}c_{k\uparrow}}\rangle|^2-|\langle{c_{-k\downarrow}c_{k\downarrow}}\rangle|^2=0,
\eeqn
where $\uparrow$ ($\rightarrow$,$\nearrow$) means that the spin is along the $z$ ($x$,$y$) direction. From Eqs.~\eqref{ScSP-1} and~\eqref{Spin-pl}, one can see that the total ScSP along both the $y$ and $z$ directions are precisely zero. This is not surprising because time-reversal symmetry is broken by the Zeeman field that is non-zero only along the $x$ direction. 

In the following discussions, we study the ScSP along the $x$ direction. The term ScSP would be used as its $x$ component if no confusion occurs. The linear dependences of $S_x(k)$ on both $k^2$ and $\epsilon_k$ result in very different behaviors of the ScSP in the low and high chemical potential limits. The numerical calculations in the rest of this work are performed using a Python package
Kwant \cite{Groth2014}. For the low chemical potential limit, represented by the lower green dashed line in Fig.~\ref{band}(b), $\epsilon_k$ is always positive, so $S_x(k)$ is positive in the whole $k$ space and antiparallel to the Zeeman field (Fig.~\ref{spin-bulk-state}(a1)]). For the high chemical potential case, represented by the higher green dashed line in Fig.~\ref{band}(b), $S_x(k)$ changes sign as $|k|$ crosses the Fermi points (Fig.~\ref{spin-bulk-state}(b1)). 

The net ScSP calculated according to Eq.~\eqref{Spin-pl} is plotted in Fig.~\ref{spin-bulk-state}(c1) as a function of $\mu$ and $M_x$. It exhibits a sharp peak at the band bottom because $S_x(k)$ is positive at each $k$ when $\mu=0$. In the high chemical potential limit, $\epsilon_k\approx\hbar v_{f}(k-k_{f})$ with $v_{f}=\hbar k_{f}/m$ in the vicinity of the Fermi surface, so $S_x(k)$ vanishes to its dependence on $(k-k_{f})$. The regime where the ScSP is most pronounced (i.e., with a sharp peak) largely coincides with the chiral regime defined by $|\mu|<M_x$ [enclosed by the green curve in Fig.~\ref{spin-bulk-state}(c1)]. This fact suggests that the ScSP could be very useful for probing topological superconductivity since it may only occur in the chiral regime. In constrast, $S_x$ decays rapidly to zero in the non-chiral regime with $|\mu|>M_x$. The evolution of $S_x$ at $M_{x}/\Delta=2$ [dashed yellow line in Fig.~\ref{spin-bulk-state}(c1)] from the chiral to non-chiral regime is plotted in Fig.~\ref{spin-bulk-state}(d1).

One can establish a physical picture for these results by analyzing two limits. For the non-chiral regime at high chemical potential, there are two spin subbands with $k_{1f}<k_{f}<k_{2f}$ [Fig~\ref{band}(b)].  For the states around $k_{2f}$ ($k_{1f}$), the $x$ components of their ScSP have positive (negative) sign [Fig~\ref{spin-bulk-state}(b1)], so they cancel each other. Despite the breaking of both time-reversal and inversion symmetries, the ScSP still vanishes in the non-chiral regime. For the chiral regime, there is only one spin subband at the Fermi surface, so the net ScSP along the Zeeman field direction remains finite. It is also noted that the ScSP slightly changes sign across the phase transition point when the system is tuned from the chiral to non-chiral regimes  [Fig.~\ref{spin-bulk-state}(d1)]. This can be explained as follows. When the chemical potential slightly increases from $\mu=0$, the ScSP for the states with $\epsilon_k < 0$ ($k<k_f$), according to Eq.~(\ref{d-vector},\ref{Spin-pl}), points along the $-x$ direction. As the density of states for a 1D system is inversely proportional to $k$, the states with $\epsilon_k < 0$ ($k<k_f$) have a larger contribution to ScSP than those with $\epsilon_k > 0$ ($k>k_f$), which results in a negative net ScSP along the $-x$ direction. 

The features of the ScSP discussed above for strictly 1D nanowire can also be found in quasi-1D systems. The multi-subband Hamiltonian for such cases is
\beqn
\label{Spinb}
H(k_x) &=& \sum\limits_{j=1}^{N} \Psi_{j,k_x}^{\dagger} \Big\{ \left[ 4t-\mu-2t\cos(k_x) \right]\tau_z \nonumber \\
	&+&2\alpha\sin(k_x)\sigma_y \tau_z+\Delta\tau_x -M_x\sigma_x \Big\} \Psi_{j,k_x} \nonumber \\
 &+&\sum\limits_{j=1}^{N-1}[\Psi_{j+1,k_x}^{\dagger}(-t-i\alpha\sigma_{x})\tau_z\Psi_{j,k_x}+\text {H.c.}]
\eeqn
where periodic (open) boundary condition is used along the $x$ ($y$) direction, $k_x$ is momentum along the $x$ direction, $N$ is the number of sites along the $y$ direction, $j$ is the site index along the $y$ direction, $t$ is the hopping amplitude and ${\Psi}_{j,k_x}=({c}_{j,k_x \uparrow}, {c}_{j,k_x \downarrow}, {c}_{j,-k_x \downarrow}^{\dag}, -{c}_{j,-k_x \uparrow}^{\dag})^T$. The results are not sensitive to the value of $N$, so we focus on the $N=2$ case below, where the Hamiltonian has two pairs of subbands with an energy splitting of about $0.7t$ due to finite size effects [Fig.~\ref{band}(c)]. We choose $t=20$meV, $\alpha=2.5$meV, $\Delta=0.9$meV, and $M_x\in [0,0.1t]$. As long as $\Delta$ and $M_x$ are much smaller than the subband energy splitting, which is exactly the experimentally relevant parameter regime~\cite{Zhang2018}, our results are not sensitive to the specific values of $\Delta$ and $M_x$.

The value of $S_{x}(k)$ at the Zeeman field $ M_x=2$meV is presented in Figs.~\ref{spin-bulk-state}(a2) and (b2) for two typical chemical potentials indicated by the green dashed lines in Fig.\ref{band}(c) (one in the chiral regime of the higher band and one in the non-chiral regime). For the former case [Fig.~\ref{spin-bulk-state}(a2)], the ScSP is positive in the whole $k$ space for the higher band, which is in the chiral regime, and changes sign with $|k|$ across the Fermi points for the lower band, which is in the non-chiral regime. For the latter case {[Fig.~\ref{spin-bulk-state}(b2)], the ScSP for both bands, which are in the non-chiral regime, change sign with $|k|$ across the respective Fermi points. The ScSP is also plotted in  Fig.~\ref{spin-bulk-state}(c2) as a function of chemical potential and Zeeman field. It is clear that the ScSP is only finite when the chemical potential is close to the band bottoms, which almost coincides with the chiral regimes enclosed by the green lines. If we fix the Zeeman energy at $M_x = 2\Delta$ and change the chemical potential along the line cut in Fig.~\ref{spin-bulk-state}(c2), the ScSP exhibits similar features as in the strictly 1D case [Fig.~\ref{spin-bulk-state}(d2)].

\subsection{Bound states}

\begin{figure*}[tb]
\centering
\includegraphics[width=2.0\columnwidth]{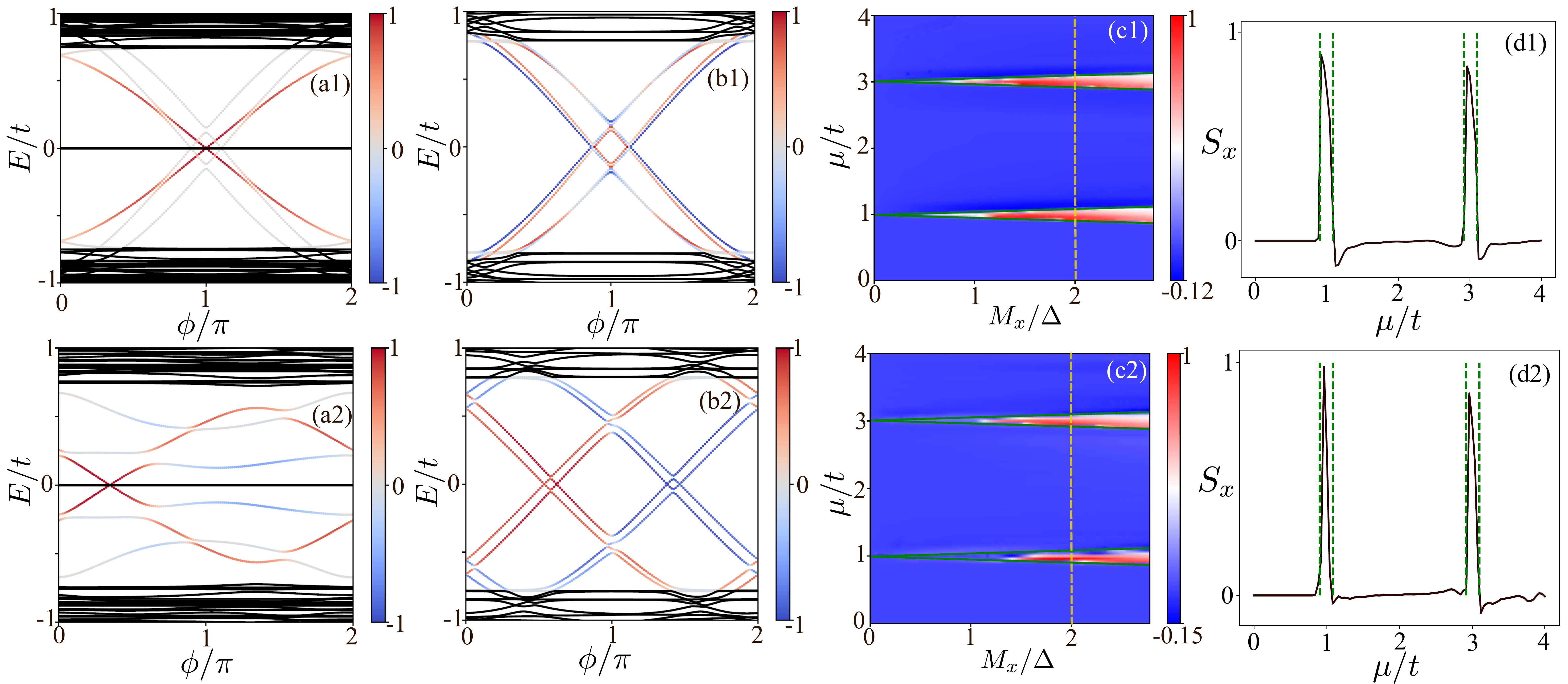}
\caption{The upper and lower panels correspond to strictly 1D and quasi-1D Majorana nanowires, respectively. The red (blue) color indicate positive (negative) ScSP. (a1,a2) E-P relation when the higher subband is non-trivial. (b1,b2) E-P relation in the trivial phase. (c1,c2) The net ScSPs of the subgap bound states as a function of the chemical potential $\mu$ and
Zeeman field $M_x$. (d1,d2) The ScSP along the vertical yellow line in (c1,c2) at Zeeman field $M_x=2\Delta$.}
\label{SNS}
\end{figure*}

In the previous section, we have studied the ScSP of the bulk states whose energies are not in the superconducting gap. As already mentioned in the introduction, the ScSP can be measured in experiments from the SOC induced anomalous Josephson current (see Sec.~\ref{Section4} for details). To this end, the ScSP of subgap states, including both trivial ABSs and MZMs, must also be understood clearly. Our results about ScAHE will be tested in two types of Josephson junctions shown in Figs.~\ref{device}(b) and (c). In both cases, a pair of quasi-1D Majorana nanowires described by Eq.~\eqref{Spinb} are connected in parallel [Fig.~\ref{device}(b)] or vertically [Fig.~\ref{device}(c)] by a normal Rashba nanowire. The length of the Majorana nanowire and normal Rashba nanowire are $L=2$ \textrm{$\mu$}m and $l=60$nm respectively. The former configuration is the most common shape of Josephson junctions and it has been proposed as a platform for measuring the $4\pi$-periodic Josephson effect due to MZMs~\cite{Kitaev2001,Fu2009,Lutchyn2010}. The latter U-shape configuration was proposed recently~\cite{Liu2016,Karzig2017} in the context of Majorana nanowires for its advantage that the Zeeman field is parallel to all superconductors. It is also a basic ingredient for building scalable Majorana-based topological quantum computers.

We first study the ScSP of the subgap states in the chiral regime. For Zeeman field $M_x=2$meV, the energy-phase (E-P) relations of the subgap states in the common-shape junction and U-shape junction are plotted in Figs.~\ref{SNS}(a1) and (a2) respectively. In both cases, the higher band is topologically non-trivial as manifested by the presence of zero energy states corresponding to the MZMs at the two far ends of the Josephson junctions as shown in Fig.~\ref{energies}(b) in appendix B. The net ScSP for the subgap states are calculated as
\beql
S_{x}(E_n)= \sum_{i,j;E_n} S_x(E_n;i,j),
\eeql
where the summation of sites $i,j$ is performed around the Josephson junction region so that we can exclude the contribution from the two MZMs at the two far ends of the device when the system is in the topological non-trivial regime (Fig.~\ref{energies}(b)). The ScSP of the MZMs,  confined in the junction and corresponding to the red curves in Fig.~\ref{SNS}(a1) and (a2), is always positive and opposite to the direction of the Zeeman field. Meanwhile, the ScSP for another two trivial subgap Andreev bound states are close to zero. In the non-chiral regime, there are four pairs of subgap states around the Fermi surface [Fig.~\ref{SNS}(b1) and (b2)]. The ScSP of the two states that are below the Fermi surface are positive (colored in red) and that of the other two states negative (colored in blue). This is consistent with what we have seen in the ScSP for the bulk states [Fig.~\ref{spin-bulk-state}(c2)]: the ScSP of the two spin subbands with $\epsilon(k)>0$ are opposite to those with $\epsilon(k)<0$.

The ScSP has also been calculated by summing over all the states below the Fermi surface. It is finite in the chiral regime (enclosed by green curves) and decay rapidly to zero in the non-chiral regime for both Josephson junctions [Fig.~\ref{SNS}(c1) and (c2)], in close analogy to what we have obtained for the bulk states [Fig.~\ref{spin-bulk-state}(c2)]. The ScSP at a fixed Zeeman field $M_x=2\Delta$ and various chemical potentials are presented in Figs.~\ref{SNS} (d1) and (d2), which are also similar to the cases of bulk states [Fig.~\ref{spin-bulk-state}(d2)].

\begin{figure}[!htb]
\centering
\includegraphics[width=0.8\columnwidth]{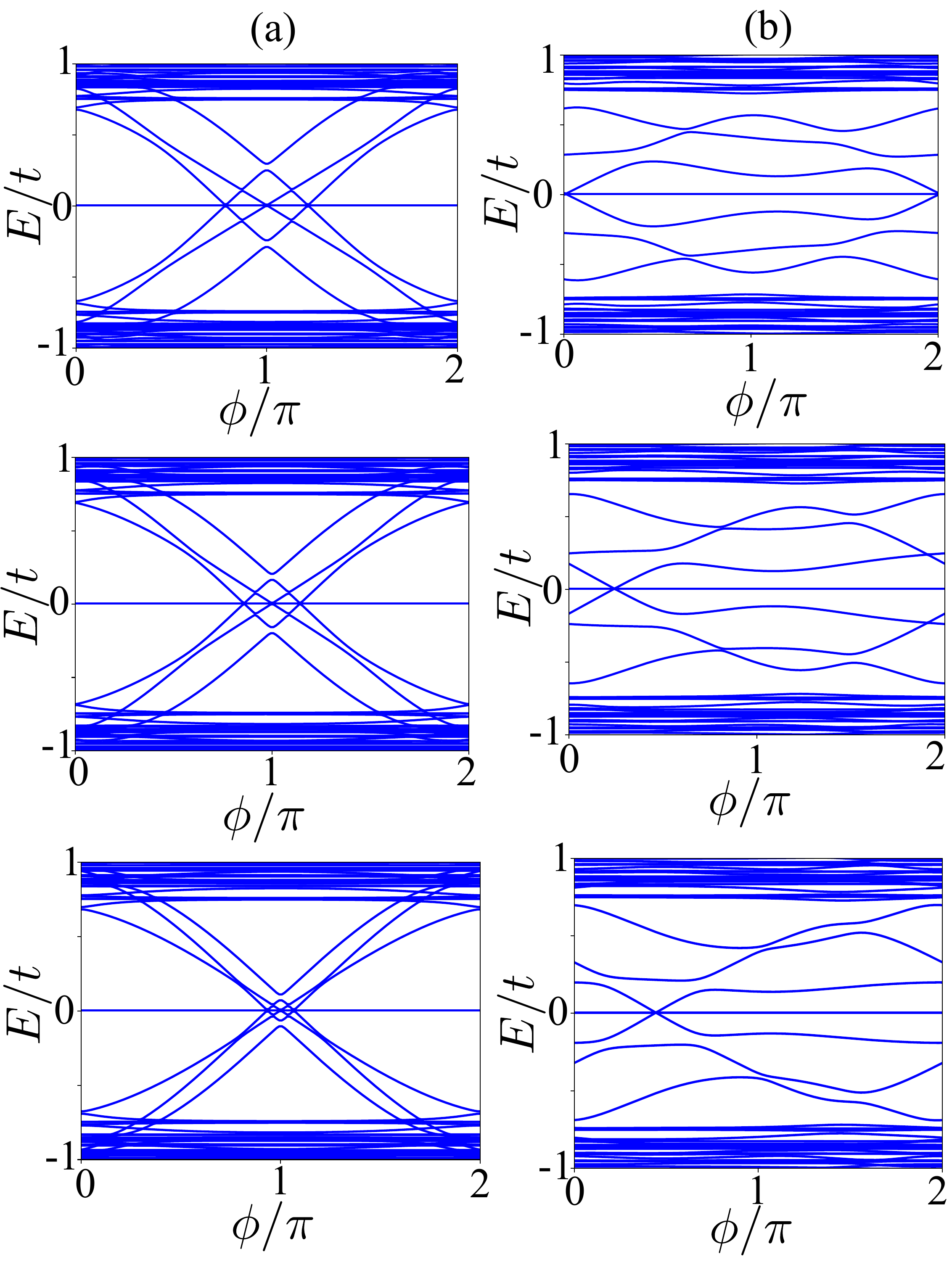}
\caption{The left and right panels are energy-phase relations in the common-shape and U-shaped Josephson junctions, respectively. The Majorana nanowire is in the topological regime. The SOC strength in the normal wire are taken to be $\alpha_n=0$, 1meV, 2meV from top to bottom.}
\label{E-P}
\end{figure}

\section{SOC-induced Josephson current}
\label{Section4}

The fact that the ScSP has a sharp peak in the chiral regime and decays very rapidly in the non-chiral regime can help us to check if a system is in the chiral regime. Based on our analysis of the ScAHE, a finite ScSP can be demonstrated using the SOC induced Josephson current. Moreover, since the SOC-induced Josephson current is perpendicular to the ScSP, it can only be observed in the U-shape junction. The energy-phase (E-P) relation of the subgap states in Fig.~\ref{SNS} already revealed this property: the E-P curves with finite slopes at $\phi=0$ only appear in Fig.~\ref{SNS}(a2), so a finite Josephson current caused by $J_s\propto\partial{E}/\partial\phi$ can only exist for U-shape junction in chiral regime.

This observation is further confirmed in the energy-phase relation for both the common-shape and the U-shape Josephson junctions (Fig.~\ref{E-P}) when the Majorana nanowire is in topological regime. Here, we take the SOC strength $\alpha_n$ in the normal wire to be 0, 1meV, 2meV which correspond to the plots from top to bottom in Fig.~\ref{E-P}. For the common-shape Josephson junction in Fig.~\ref{E-P} (a), the SOC only shifts the E-P relation of the trivial ABSs but has negligible effect on the MZMs. The E-P curves of the MZMs all cross at $\phi=\pi$ and the slopes of all curves at $\phi = 0$ [since we calculate the current without flux] are zero regardless of the SOC strength. One concludes that the Josephson current is always zero at $\phi=0$. On the contrary, the E-P curves of the U-shape junction in Fig.~\ref{E-P}(b) exhibit very different features even though the ScSP of both Josephson junctions are similar. Firstly, the E-P curves of the MZMs confined in the junction cross at $\phi=0$ when the SOC strength is zero. It is consistent with our previous study about strictly 1D systems~\cite{Liu2016} that the Josephson coupling in the common-shape and U-shape junctions have a $\pi$ phase shift. Secondly, the E-P curves for both MZMs and trivial ABSs are shifted as the SOC strength increases, which implies that there is an SOC induced Josephson current at $\phi=0$. This Josephson current is along the $y$ direction and perpendicular to the ScSP, which agrees with our analysis of the ScAHE and in consistence with our previous work for strictly 1D systems \cite{Liu2016}

\begin{figure}[!htb]
\centering
\includegraphics[width=1.0\columnwidth]{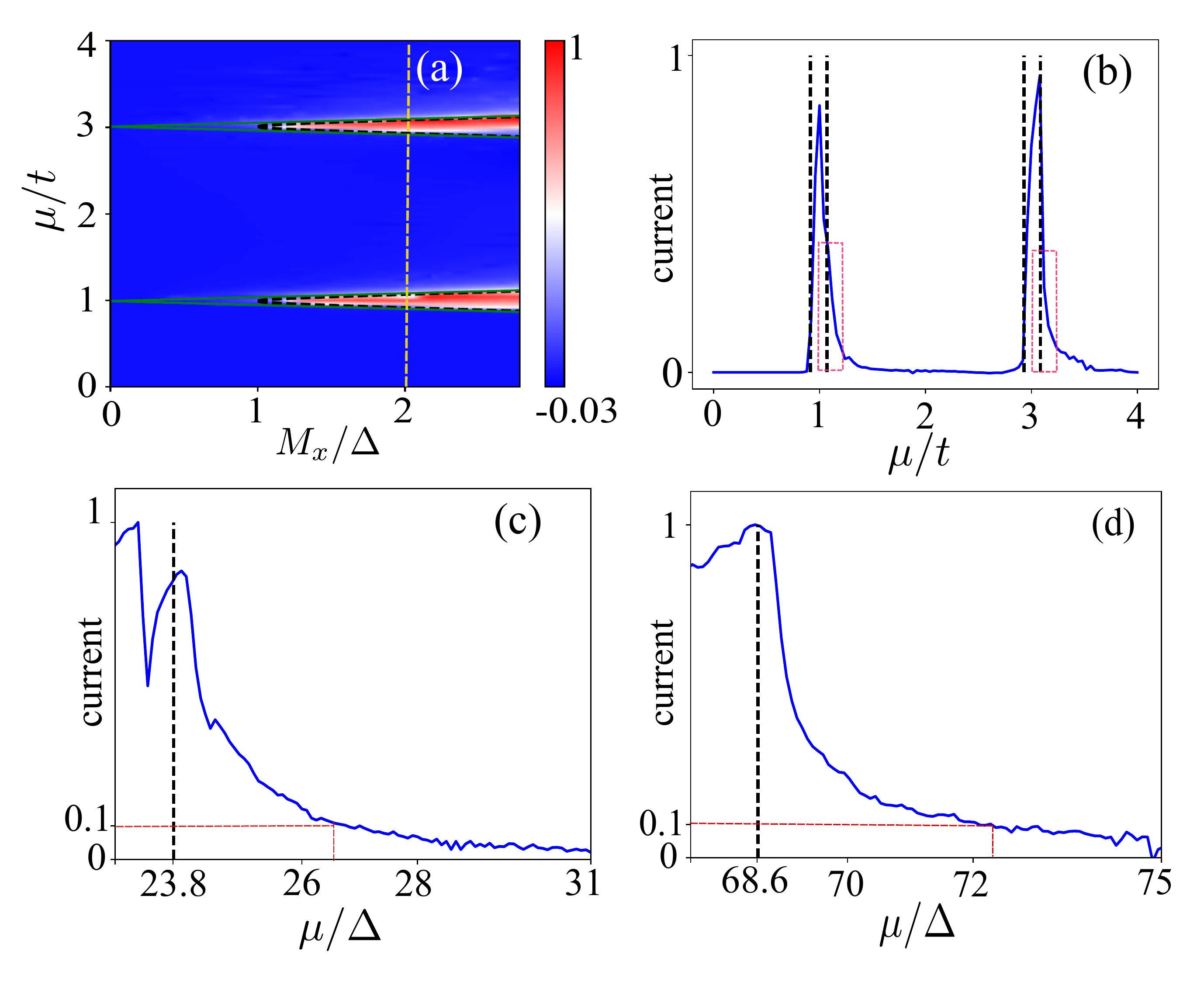}
\caption{(a) The Josephson current in the U-shaped jucntion as a function of the chemical potential $\mu$ and the Zeeman field $M_x$. The green lines indicate the boundary of the chiral regime and the black dashed lines indicate the boundary of the topological regime (see also Appendix B). (b) The Josephson current along the vertical yellow dashed line in (a) at a fixed Zeeman field $M_x=2\Delta$. (c,d) The decaying behaviors of the Josephson current in the regime enclosed by the two red rectangles in (b). The right side of the dashed black line is the trivial regime.}
\label{JC}
\end{figure}

For the U-shape junction at $\phi=0$ and $\alpha_n=1.5$meV, the Josephson current as a function of chemical potential and Zeeman field is presented in Fig.~\ref{JC}(a). The Josephson current is not always finite in the entire chiral regime (enclosed by the green solid curves): it is finite in the topological non-trivial regime but almost zero in the topologically trivial chiral regime with the detailed definition in appendix B. This is because the Josephson current is more sensitive to boundary conditions in the trivial regime, and may be suppressed more easily. The Josephson current is plotted in Fig.~\ref{JC}(b) as a function of $\mu$ along the $M_x=2\Delta$ line cut in Fig.~\ref{JC}(a). While the current remains finite in a very narrow region in the vicinity of the phase boundary on the trivial side, its magnitude reduces to less than 10$\%$ as the chemical potential deviates from the phase boundary by less than $4\Delta$ [Fig.~\ref{JC}(c) and (d)]. It is very likely that the observation of a sharp peak in the SOC-induced Josephson current would help us to locate the chiral or even the topological regime.

\section{Discussion and conclusion}
\label{Section5}

This work studies the ScSP and related anomalous Josephson current, which remains finite for uniform superconducting phase, in quasi-1D superconducting nanowires.  As the ScSP driven anomalous Josephson current only happens in the U-shape junction within the regime almost coincident with topological regime, it can provide an characteristic signals of MZMs related to their spin degree of freedom. We note that the Josephson junction with finite Josephson current at zero phase difference, which is refereed as $\varphi_0$-junction, has bee studied in common-shape Josephson junctions~\cite{Buzdin2003,Buzdin2008,Szombati2016,Schrade2017}. In these previous works, some proposals require stringent conditions such as non-uniform magnetic field~\cite{Buzdin2003} or conventional superconductors with totally different Zeeman field directions~\cite{Buzdin2008}. The U-shape Josephson junction is advantageous because the ScAHE effect can only be observed in such systems with the Zeeman field parallel to all Majorana nanowires. It is also a basic element for building scalable topological quantum computers~\cite{Karzig2017}, so we believe that our results are of general interest.

In conclusion, we have analyzed the general properties of the ScSP in superconductors and investigated its specific behaviors in 1D nanowires, with the contributions from bulk states, trivial ABSs, and MZMs identified separately. For two types of Josephson junctions, the ScSP in all three cases exhibits a very sharp peak in the chiral regime and rapidly decays to zero in the non-chiral regime. For a superconducting thin film with in-plane spin polarization and an out-of-plane electric field, we uncover an important phenomenon called the ScAHE. An SOC induced supercurrent can be observed in the U-shaped Josephson junction made from 1D nanowires with non-zero ScSP. This occurs almost concurrently with the ScAHE, which has a sharp peak in the topological regime but becomes negliglible in the trivial regime. This work demonstrates that the spin properties of MZMs lead to special Josephson effect that would facilitate their experimental detections. It is a reliable method that can differentiate MZMs from bulk states and trivial ABSs because their contributions are well-understood and clearly different.

\section*{Acknowledgement}

We would like to thank Dong-Ling Deng, Chun-Xiao Liu, Xiong-Jun Liu, and Hong-Qi Xu for useful discussions. This work is supported by National Key R\&D Program of China (Grant No. 2016YFA0401003),  NSFC (Grant No.11674114), Thousand-Young-Talent program of China, and the startup grant of HUST.

\appendix

%\begin{widetext}

\section{Rashba SOC induced anomalous kinetic momenta for spin-triplet pairings}

In this section we provide the details on the derivation of the ScAHE. In general, a wave function without SOC in the Nambu space has the form
\begin{eqnarray}\label{wf-10}
\psi(\lambda=0)=\left(\begin{array}{c} \psi_{e} \\ i\sigma_y \psi_{h} \end{array}\right),
\end{eqnarray}
where $\psi_{e(h)}$ is the spinor wave function for the electron (hole). When we add Rashba SOC ($\lambda \bm{\sigma} \times \nabla V$) in the system, the electron wave functions are transformed as
\begin{eqnarray}\label{1}
\psi_{e}(\lambda) &=& \exp \left[ -\frac{i}{\hbar} \int_0^{\bm r} \lambda (\bm{\sigma} \times \bm{\nabla} V)  \cdot \bm{dr} \right] \psi_{e}(\lambda=0) \nonumber \\
 &=& \hat{U}_{e}(\lambda)\psi_{e}(\lambda=0).
\end{eqnarray}
Similarly, the spinnor wave function for the hole will be transformed as
\begin{eqnarray}\label{2}
\psi_{h}(\lambda) &=& \exp \left[ \frac{i}{\hbar} \int_0^{\bm r} \lambda (\bm{\sigma}^* \times \bm{\nabla} V)  \cdot \bm{dr} \right] \psi_{h}(\lambda=0) \nonumber \\ 
 &=& \hat{U}_{e}^*(\lambda)\psi_{h}(\lambda=0).
\end{eqnarray}
Thus the wave function is transformed as
\begin{eqnarray} \label{3}
\psi(\lambda)&=&\left[ \begin{array}{c} 
\hat{U}_{e}(\lambda)\psi_{e}(\lambda=0) \\ 
i \sigma_y \hat{U}_{e}^*(\lambda)\psi_{h}(\lambda=0) 
\end{array} \right] \nonumber \\
& = & \left[ \begin{array}{c} 
\hat{U}_{e}(\lambda)\psi_{e}(\lambda=0) \\
\hat{U}_{e}(\lambda) i \sigma_y \psi_{h}(\lambda=0) 
\end{array} \right] = 
\hat{U}(\lambda)\psi(\lambda=0). \nonumber
\end{eqnarray}

For a superconducting thin film, an out-of-plane electric field generates a Rashba SOC term
\beql
\hat{H}_{so}=\lambda E_z(p_x\sigma_y-p_y\sigma_x)\tau_z.
\eeql
The gauge transformation matrix for the electron wave fucntion becomes $\hat{U}_{e}(\lambda)=\exp\left[-\frac{i}{\hbar}{\lambda}E_z(x\sigma_y-y\sigma_x)\right]$, so the superconducting condensates are transformed as
\begin{eqnarray}\label{4}
\hat{\Phi}(\lambda) &=& \hat{U}_e \hat{\Phi}(\lambda=0)\hat{U}_{e}^{\dagger}.
\end{eqnarray}
This does not change spin-singlet pairing, but modifies the spin-triplet condensation as 
\begin{eqnarray}
\bm{d}^{\lambda} \cdot \bm{\sigma} = \hat{U}_e \bm{d}^{0} \cdot \bm{\sigma} \hat{U}_{e}^{\dagger}
\end{eqnarray}
For $d_{x}^{0}\sigma_x$, expanded to the first order of $E_z$, we obtain
\begin{eqnarray}\label{5}
d_{x}^{\lambda}\sigma_x &\approx& \left[ 1-\frac{i}{\hbar}\lambda E_z(x\sigma_y-y\sigma_x)\right] d_{x}^{0}\sigma_x \nonumber \\
&\phantom{=}& \times\left[ 1+\frac{i}{\hbar}\lambda E_z(x\sigma_y-y\sigma_x) \right] \nonumber \\ 
&\approx& \exp\left(-2i\lambda E_{z}x\sigma_y\right) d_x^0 \sigma_x. \nonumber
\end{eqnarray}
In general, we have 
\beqn \label{6}
\bm{d}^{\lambda} \cdot \bm{\sigma} &\approx& \exp \left[ -\frac{i}{\hbar}2\lambda E_z x \sigma_y \right] d_x^0 \sigma_x \nonumber \\
&& + \exp \left[ \frac{i}{\hbar}2\lambda E_z y \sigma_x \right] d_y^0 \sigma_y \nonumber \\
&&+  \exp \left[ -\frac{i}{\hbar}2\lambda E_z (x\sigma_y - y \sigma_x) \right] d_z^0 \sigma_z,
\eeqn
which lead to the additional spin-dependent term for the kinetic momenta shown in Table.~\ref{table:table1}.

\section{The topological and chiral regimes}

In this section, we define the topological and chiral regimes. For the strictly 1D case, we know that the topological phase boundary is determined by $\Delta^2+\mu^2<M_x^2$. Here we found that for multi-band cases with the parameters of our interest, the regime defined as $\Delta^2+(\mu-\mu_i)^2<M_x^2$, enclosed by black dashed curves in Fig.~\ref{JC}(a) are topological phase boundary. $\mu_i$ indicated by green dash lines in Fig.\ref{energies}), is the chemical potential of $i$-th band at $k=0$ without magnetic field. In Fig.~\ref{energies}(b), we show the lowest positive eigenenergy $E_g$ of the quasi-1D Majorana nanowire with $N=2$ as a function of the chemical potential $\mu$ and the Zeeman field $M_x$. We found the regimes with zero eigenenergy precisely coincide with the the regime $\Delta^2+(\mu-\mu_i)^2<M_x^2$ indicated by the green curves. The chiral regime is defined by $|\mu-\mu_i|<M_x$ which contains the topological regime. The trivial chiral regime satisfies $|\mu-\mu_i|<M_x<\sqrt{\Delta^2+(\mu-\mu_i)^2}$. 

\begin{figure}[htb]
\includegraphics[width=1.0\columnwidth]{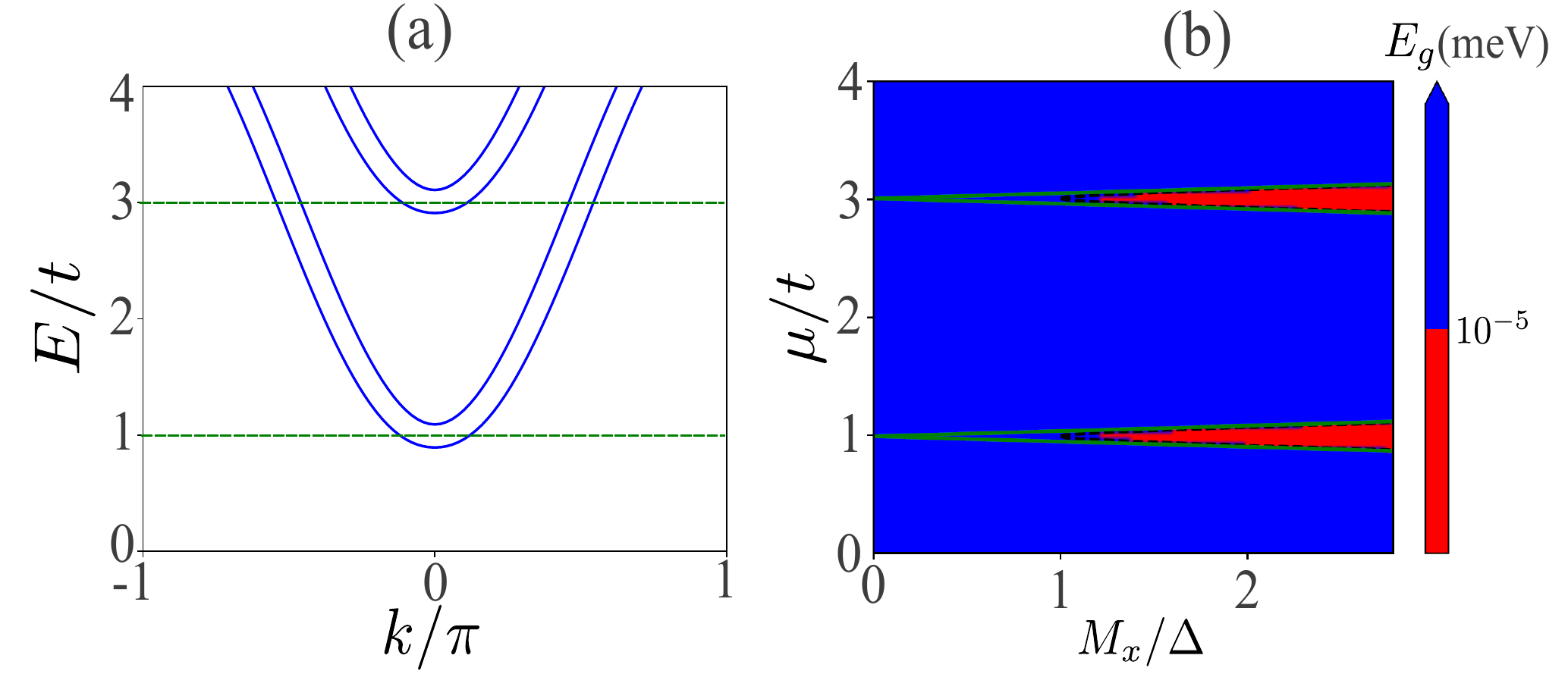}
\caption{(a) The electron dispersion of nanowire  (b) Lowest eigen-energy of the nanowire as a function of chemical potential and Zeeman field. The red and blue region indicates the nontrivial and trivial region which is determined by the existence of zero energy. The black dashed curves enclose the regime with $ \Delta^2+(\mu-\mu_i)^2<M_x^2$ where $\mu_i$ is the chemical potential at $k=0$ without Zeeman field for the $i$-th band as indicated by the green dash lines. The green curves enclose the chiral regime satisfying $|\mu-\mu_i|<M_x$.}
\label{energies}
\end{figure}

%\bibliography{Topo-phy-1}
%

\end{document}